2-16-07

# Dirt, Gravity, and Lunar-Based Telescopes: The Value Proposition for Astronomy

DAN LESTER[1]
[1]Department of Astronomy C1400, University of Texas, Austin TX 78712

## Abstract

The lunar surface has historically been considered an optimal site for a broad range of astronomical telescopes. That assumption, which has come to be somewhat reflexive, is critically examined in this paper and found to be poorly substantiated. The value of the lunar surface for astronomy may be broadly compelling only in comparison to terrestrial sites. It is suggested here that the development and successful operation of the Hubble Space Telescope marked a turning point in the perception of value for free-space siting of astronomical telescopes, and for telescopes on the Moon. As the astronomical community considers the scientific potential of the *Vision for Space Exploration* (VSE) and the return to the Moon in particular, it should construct a value proposition that includes the tools, technology, and architecture being developed for this return, as these can well be seen as being more astronomically enabling than the lunar surface itself – a destination that offers little more than rocks and gravity. While rocks and gravity may offer astronomical opportunity in certain scientific niches, our attention should be focused on the striking potential of human and robotic dexterity across cis-lunar space. It is this command of our environs that the VSE truly offers us.

---

## 1. Destination and Architecture

With the implementation of the President's *Vision for Space Exploration*, and the planned return by human beings to the Moon in the next decade, the science community has been offered new capabilities, and a chance to develop accompanying science programs for them. For the astronomical community, the lunar surface has historically appeared promising as a future site for highly capable telescopes. We have risen to this challenge before, notably with the AIP Annapolis workshop on "Astrophysics from the Moon" (1990) in response to the Space Exploration Initiative proposed in 1989. Also relevant in this context is the AIP Stanford "Physics and Astrophysics from a Lunar Base" (1989) workshop, organized in the run-up to the SEI announcement. These workshops identified what were, at the time, opportunities offered *only* by observatories on the Moon.

In the decade and a half that passed since then, our perspective on astronomical instruments in space changed dramatically. We have now witnessed the completion and successful operation of all four Great Observatories: Hubble Space Telescope, Compton Gamma Ray Observatory, Chandra X-Ray Observatory, and the Spitzer Space Telescope. These Great Observatories, deployed in free-space, represent achievement of a high level of confidence in civil-space optical and sensor systems engineering, as well as precision pointing and tracking. These accomplishments, along with a host of smaller scale free-space scientific missions, throw entirely new light on the future of astronomical telescopes in space. In many respects, these accomplishments were built on the understanding gained from, if not the specific architecture of, the lunar exploration program that preceded them by several decades. Most of the scientific thrusts proposed for the lunar surface in those early workshops have, as it turned out, either been achieved in free space, or are anticipated with what are now technologically credible free-space designs.

It is with this understanding that the organizers of the present workshop have wisely recognized that the value to astronomy from Exploration is more than a specific destination. The name of this workshop – "Astrophysics Enabled by the Return to the Moon" reflects that explicitly. We are going back to the Moon, and are developing the capabilities to do so. Scientists who study things in space other than the Moon itself need to look at those capabilities as having broader application than for studies of that particular piece of rock in space. The *Vision for Space Exploration* was conceived to lead us back beyond low Earth orbit and, in the process, to enable human travel in cis-lunar space as well as building on our earlier achievements with humans on the lunar surface. As currently being implemented, the basis for this will be the *Orion/*Crew Exploration Vehicle (CEV which, along with the Lunar Surface Access Module (LSAM), will bring human dexterity, expertise, and intelligence to build, deploy, and maintain scientific facilities throughout cis-lunar space. The *Ares 5* heavy lift launcher will also provide opportunities for lofting astronomical telescopes and components for those telescopes that far exceed the light gathering power and complexity of facilities we can now envision. New robotic technology will allow relatively simple maintenance tasks on telescopes to be performed either autonomously or telerobotically. When combined with humans in free-space, this robotic technology will dramatically extend their reach, multiplying their efficiency and enhancing their skills. Much more than realizing an ability to use the lunar surface as an observatory site, the evolving Exploration Architecture offers the potential to put humans firmly into space astronomy, as they're hands-on involvement is so powerful for ground-based astronomy. This capability can be built upon the regular servicing of the Hubble Space Telescope, which proved that human beings were enabling for astronomical accomplishment in space.

For astronomers, the *Vision for Space Exploration* has to be seen as providing enabling architecture, rather than directing us to a particular destination. More broadly, in which destinations can include places in free-space such as Lagrange point orbits as well as on the surfaces of massive bodies, the scientific value of the Exploration Architecture becomes especially exciting. The question then becomes to what extent the lunar surface itself is actually enabling for astronomy. This is a strategic question that has to be answered by our astronomical community as we look at our scientific priorities and available technology. Our effort on this question, with respect to the *Vision for Space Exploration*, has been lacking, however, and this workshop should provoke serious needed thought.

## 2. What the Moon Was to Astronomy

*"So many factors favor the Moon as a site for future large-scale space astronomy that planning an observatory there deserves the closest attention in the years just ahead."*

*William Tifft*
*University of Arizona*
*Aeronautics and Astronautics, December 1966*

In considering the importance of the lunar surface to astronomy, it is helpful to examine the reasons why it was originally so highly valued. A careful assessment of this value was developed by Tifft (1966), in which the technology of the era clearly pointed to the lunar surface as an enabling place for astronomical telescopes. In that era, astronomical sensing was done entirely with photographic emulsions and photomultipliers. The capabilities of pointing/tracking for telescopes in space at that time was exemplified by OAO-2 (the first real observatory in space, launched in 1968), for which 1'/1" pointing/tracking capabilities were not a lot better than that achievable with terrestrial telescopes. In the case of OAO-2, tracking was done with a quadrant detector and strapdown gyros, but it was realized that such stabilization would be particularly difficult in a space-station environment where human attendants and observers (needed at least for chaning photographic plates!) were moving about in the cabin. In this context, the Moon was a large reaction mass that, by anchoring the telescope to it, allowed for efficient decoupling of astronauts from the pointing system. It was the largest natural reaction mass nearby that had no obscuring atmosphere, offering both a panchromatic perspective and diffraction-limited performance at all wavelengths. On the Moon, one could use what were seen, at that time, cutting edge and proven ground-based astronomy tracking technologies to follow the slow motion of the sky overhead, with the added advantage that the surface of the Moon is seismically quiet compared to the Earth.

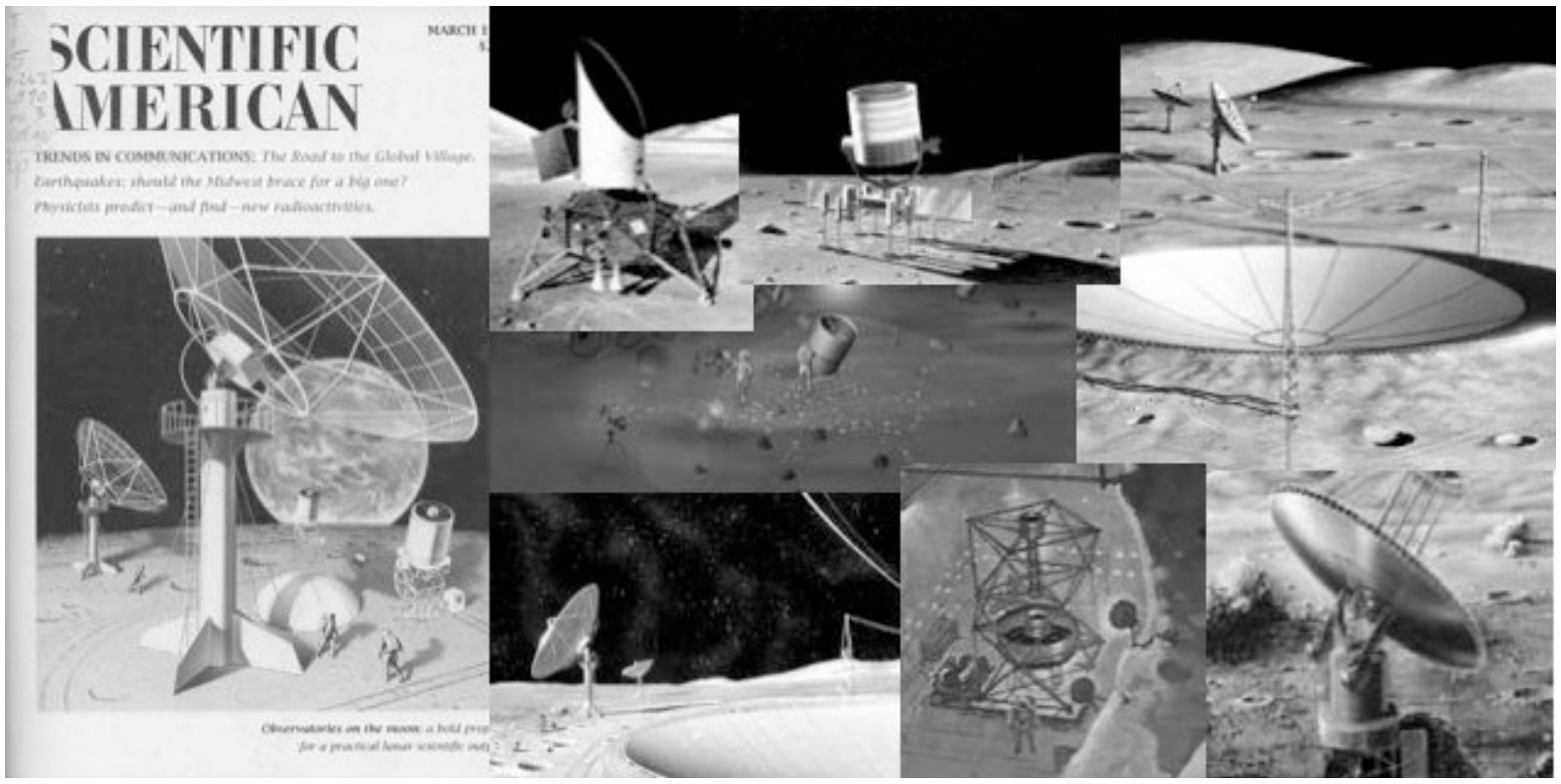

Figure 1: This history of lunar astronomy is a rich one, based on the idea that telescopes need to be planted on something in order to work properly. An enormous amount of creativity and innovation have been expended on this idea, with efforts peaking in the early 1990s, before the development of HST, and loosely coupled with the Space Exploration Initiative.

Although the challenges of telescope stabilization with humans onboard could not be met (and perhaps still cannot be met!), Tifft’s conclusions about the value of humans for astronomical observations – bringing skill, versatility, and intelligence to what might be complex equipment requiring real-time decision making are, in the context of technological improvements we’ve now realized, strikingly appropriate even today. Tifft articulated this in the unfortunately gender-specific language of the time. *“Man, despite his disturbing influences, represents a versatile, adaptable, and reliable, ready-made system which simply cannot be artificially duplicated. In his simplest form, man can assemble or deploy complex systems, he can adjust and maintain them, and he can periodically update them by changing accessories or making additions or modifications”* These visionary points could be used today to describe the success of the construction of the International Space Station and servicing of HST.

It should be noted that astronomy has been done from the surface of the Moon. The Apollo 16 far-UV (<1600Å) camera/spectrograph -- UVC (Carruthers 1973) was a 22 kg package, deployed by the astronauts (with a copy used later on Skylab). This was a manually pointed instrument that was left on the lunar surface and, with a field of view of ~20º, this 3-inch aperture electronographic Schmidt camera was hardly a telescope at all. As a result, it needed no active tracking. In the present terminology we would consider this an example of “suitcase” lunar astronomy, where a self-contained, low mass instrument could be stowed on an auxiliary supply pallet. The UVC could be set up and operated by an astronaut, with appropriate attention to contamination mitigation from the portable life support system. The instrument depended on the astronaut to deploy it in the shadow of the LEM, level it by adjusting the legs, attach it to a separate battery module with cables, point it manually using a sighting tool and graduated circles, and then trigger the shutter. In addition to images of the Earth’s geocorona, this instrument was used to map the UV radiation from the Magellanic Clouds in order to assess the contribution from hot, young stars there.

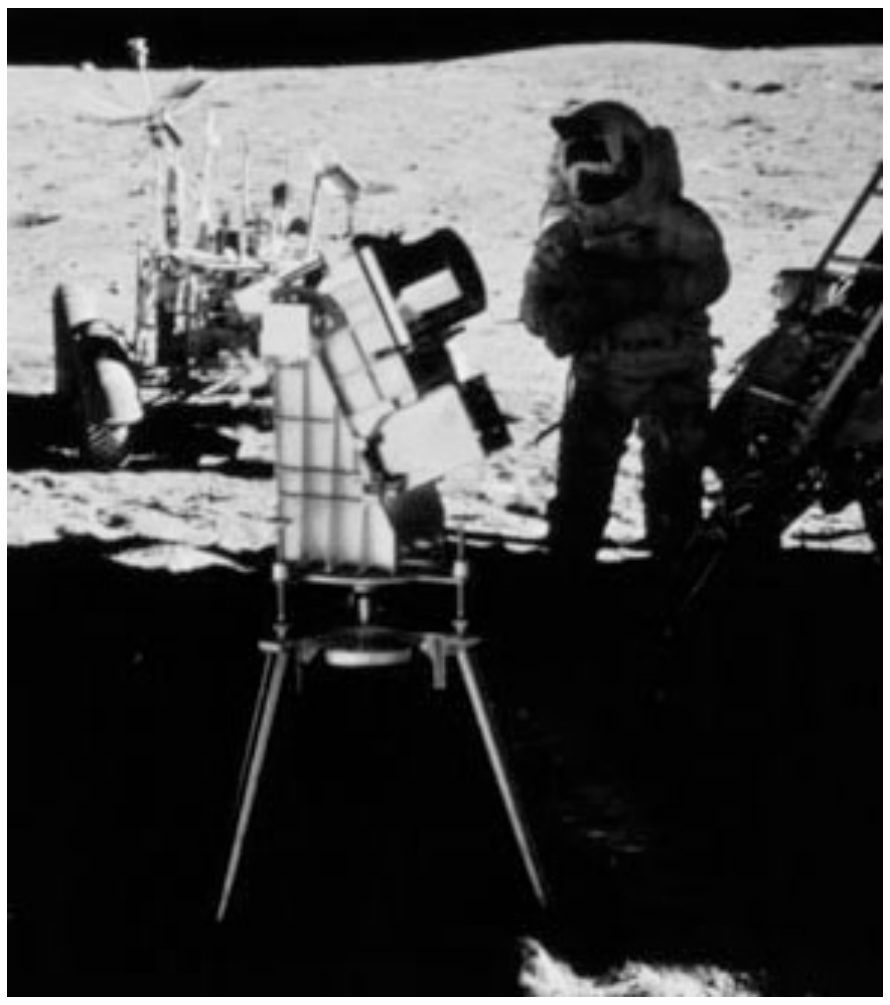

Figure 2: Apollo 16 astronaut John Young stands next to the Far Ultraviolet Camera/Spectrograph, a 3-inch telescope recording electronographically on returnable emulsion at 500-1600Å. This represents the only astronomical instrument ever deployed on the surface of the Moon. As a suitcase-sized instrument *in vacuo*, in an era in which free-space tracking and guiding was primitive, and in which data return required physical reentry into the Earth’s atmosphere and subsequent recovery, this was an enabling science instrument.

With high bandwidth communication channels that we have today, and reliable and efficient electro-optical and electro-mechanical servo systems, the operation of astronomical observatories is no longer a hands-on proposition, either on Earth or in space. But the availability of human agents for intervention (planned or unplanned) is in principle advantageous for just about any science experiment. In some scenarios, a lunar surface telescope is sited near a permanently occupied base. In this case, visits by astronauts might well be frequent, offering some benefit for instruments needing more intensive care and attention.

The lunar surface was also recognized as offering unique thermal environments. Thermal stability is a serious optical alignment issue at short wavelengths, and frequent solar eclipses makes low inclination low Earth orbits particularly challenging in this regard. While most of the lunar surface endures large temperature swings, the timescale is days rather than minutes. In the case of permanently shadowed lunar polar craters, thermal equilibrium is almost trivial, albeit at low temperature, with natural heat loads determined by scattered sunlight, and a miniscule lunar heat flow.

On the basis of the Apollo-deployed seismometers, it was understood that the lunar surface was significantly quieter than the surface of the Earth and, as a result, telescopes mounted on the Moon would suffer less seismic-induced misalignment than ones on the Earth. Without tectonic activity, it is understood that the residual natural seismic activity is most likely the result of impacts. As a naturally quiet physical surface, the Moon was considered to offer optical-bench like properties that could enable linked telescopes for interferometry. Because of the large reaction mass provided by the Moon, and the slow motion of the celestial sphere over it with no atmosphere, terrestrial telescope technology could offer exceptionally precise pointing and tracking.

It was realized rather early in the consideration of the lunar surface for astronomy (Gorgolewski 1965) that the Moon can provide very effective shielding of radio frequency noise from terrestrial sources, both auroral and manmade. As such, the far side of the Moon offers at least an order of magnitude lower RF background than the near side. As a result of the substantial size of the Earth's geotail, and also as a result of diffraction of that emission around the limb, radio telescopes down on the far side surface (ideally in a crater there) are especially promising in this regard. While the Moon does have a very tenuous ionosphere, the plasma frequency of which is not yet known, it is certainly the case that it is much lower than that of the Earth, offering transmission well below 10MHz. The lunar surface offers, in this respect, a functional backplane that allows a wire grid antenna to be monodirectional.

On the basis of these characteristics of the lunar surface, built on the advantages of no obscuring and distorting atmosphere, and in view of the difficulties of operating in free-space, many efforts were understandably and justifiably directed towards refining concepts for lunar surface astronomy. Early U.S. efforts were detailed in the definitive collections referred to above and well summarized by Burns et al. (1990). This interest in the astronomical community drove engineering efforts – aerospace, mechanical, and even civil – to refine concepts for how telescopes could be put on the lunar surface (Johnson et al. 1990, Van Susante 2002, Duke & Mendell 2002). This has been supported with expertise by lunar science authorities who bring wisdom about terrain and geology to bear (e.g. Lowman 1995). The concepts thus developed were creative and innovative. Given reasonable projections of lunar surface capabilities based on our Apollo experience, installation and operation of telescopes on the lunar surface appeared credible.

A critical element of these lunar astronomy ideas was the presumption that such telescopes were, in the context of future human basing on the lunar surface, routinely accessible, whether or not human beings were stationed nearby. The idea that human beings stationed there could walk over to maintain and service – perhaps even construct or deploy – an astronomical telescope was entirely consistent with the paradigm of terrestrial astronomy, where the best facilities were enabled by extensive local hands-on engineering support.

As a result of these considerations, a number of different explicit concepts for lunar telescopes were proposed by astronomers, of more variety than that originally envisioned by Tifft. Several of these are represented at this workshop. A VLF interferometer based on a large array of wire grids was first proposed by Douglas and Smith (1985). A UV-optical transit telescope that made use of the slow motion of the sky across the Moon was first proposed by McGraw (1994) and further developed by team members Nein and Hilchey (1995). Shorter wavelength interferometric capabilities on the Moon were suggested by Labeyrie (1993), taking advantage of the baseline hosting opportunities provided by the lunar surface. Modest sized UVOIR fully steerable robotic telescopes, depending on novel, highly lightweighted optics have been proposed as well (Chen et al. 1995). Many more recent conceptual studies have concentrated on these general ideas. Most recently, a unique innovative concept to deploy a large aperture liquid mirror telescope on the Moon has been proposed (see below, Angel et al. 2005, 2006).

## 3. How HST Diminished the Astronomical Promise of the Moon

The development of the Hubble Space Telescope marked a fundamental change in the outlook for lunar surface astronomy. Although I don't believe it has been explicitly noted, following the successful COSTAR servicing of Hubble and proof of the originally planned capabilities for that telescope in low Earth orbit, the enthusiasm for lunar surface telescopes as measured by the number of papers devoted to them dropped substantially. HST fundamentally altered the way we look at space astronomy, and with it we fully demonstrated free-space capabilities that negated almost all of the perceived uniqueness of the lunar surface.

For the first time, with HST, we had widefield diffraction-limited performance with no atmospheric obscuration. In doing it, we had proved our ability to handle an extraordinarily tough thermal environment, with sunrises and sunsets every ninety minutes. Without a stable surface to mount the telescope on, we achieved routine pointing and tracking at a level of accuracy (0.003") never before achieved on the surface of the Earth. This was done with high observational efficiency, and in the course of more than a decade in orbit, demonstrated survivability in space over a long period of time.

The high data rate needed for large format imagers was easily achieved, as was largely autonomous operation. Most importantly, as proven in the course of four successful servicing missions over its lifetime to date, we demonstrated hands-on accessibility of HST for servicing and maintenance, allowing us to realize space performance with ground-based reconfigurability. We now look back at these capabilities for space astronomy derived in many cases from technology that is about twenty to thirty years old. With the technological achievements that followed the development of HST, and particularly the tremendous advances in space construction and servicing proven on the International Space Station, we can make credible extrapolations of these free-space capabilities as we consider how the evolving Exploration Architecture may enable even more ambitious space astronomy missions.

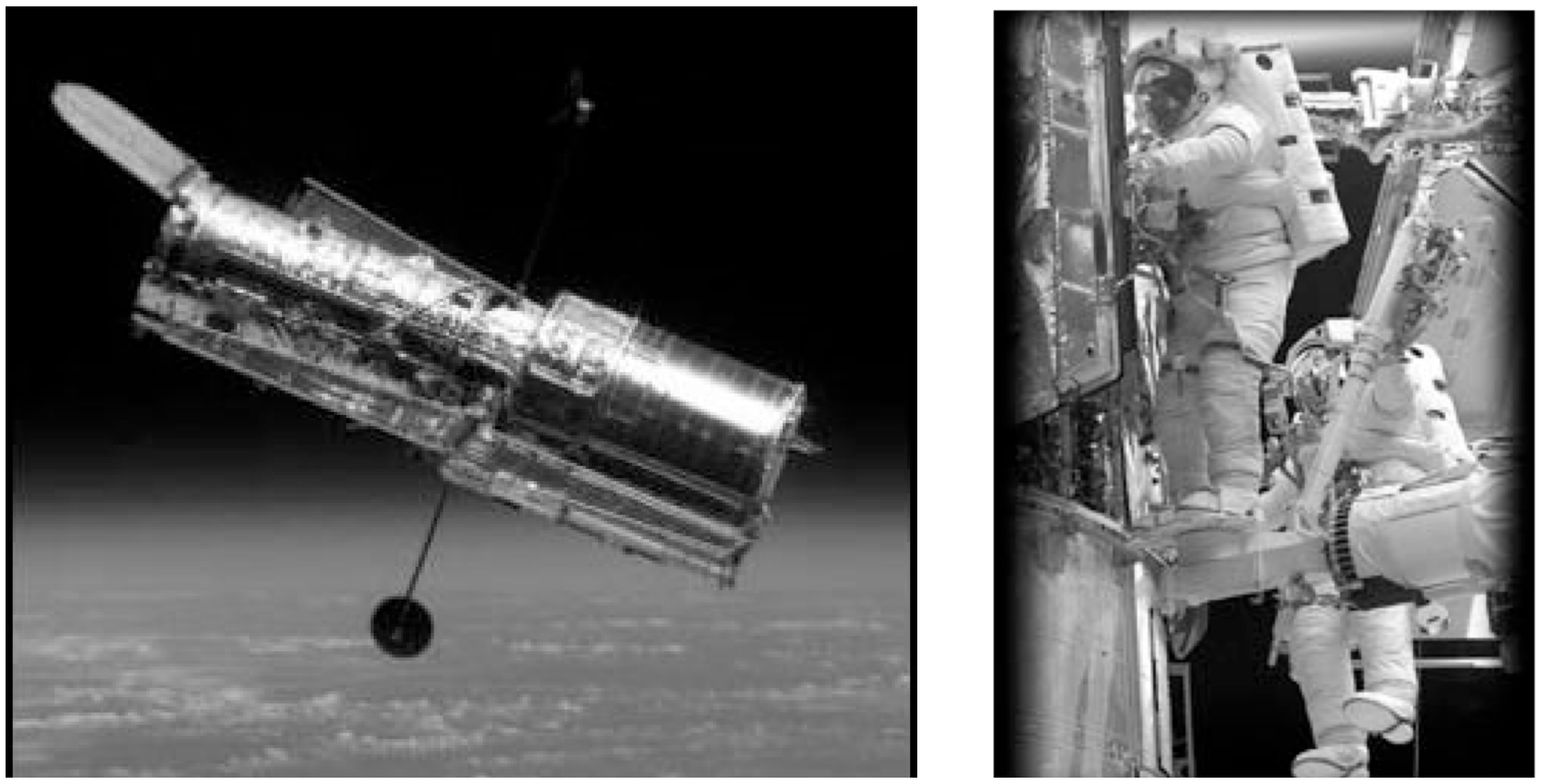

Figure 3: The successful operation of the Hubble Space Telescope changed the paradigm about free space as a powerful site for optical and ultraviolet astronomy. The astronaut servicing of HST has repeatedly demonstrated that, with regard to astronomical instrumentation, hands-on reconfiguration and repair is by no means a capability that might be viewed as unique to the lunar surface.

Largely as a result of our success with HST, a number of the most important reasons for putting at least a single dish UV/optical telescope on the surface of the Moon essentially evaporated. With the announcement of the *Vision for Space Exploration*, and new plans to return to the Moon, this position was argued in some detail by Lester et al. (2004). A recent discussion by Lowman and Lester (2006) summarizes this, as well as a contrasting argument.

## 4. Lunar Astronomy in the Context of Space Science Strategic Planning

Now that our country has made, with the *Vision for Space Exploration*, the commitment to return to the Moon and develop our capabilities for at least cis-lunar space, it behooves the astronomical community to look carefully for opportunities that commitment might offer. This is, as we understand, a key goal of this workshop. While science has been declared as <u>not</u> being a fundamental driver for the VSE, it is probably simplistic to reflexively conclude that VSE does not offer enabling opportunities to astronomy. The relationship of astronomy to lunar exploration needs to be evaluated carefully, and couched in a strategic context, especially in view of our hard won capabilities in free-space.

The first question we need to ask about astronomy and lunar exploration is whether we can, in fact, deploy and service telescopes on the lunar surface. As noted above, there is a rich history of innovative and creative planning that says that we probably can, though at a cost burden that is not well understood. This involves not just getting big pieces to the lunar surface, but assembling telescopes out of them and maintaining them there. Much of the last thirty years of lunar telescope concept development ends at this point, presuming that siting of telescopes on the lunar surface is optimal as well as possible. I believe this presumption is misguided.

In this context, a second question is whether the lunar surface offers uniquely enabling opportunities for priority astronomical research. As discussed above, I would say that in many respects it used to. From an engineering standpoint, compared with cis-lunar free space, the lunar surface offers two unique characteristics unambiguously – gravity (and concomitant reaction mass) and rocks (grit, dust, and regolith). While the current implementation plan for the VSE may add to these the possibility of continual human presence near a space telescope, and future ISRU developments might provide a resource base for astronomical hardware, neither of these two unique characteristics are conspicuously enabling for the astronomical instruments that we have thus far considered important. There is no question that lunar siting is vastly better – albeit vastly more expensive – for astronomy than terrestrial siting, but free space siting is vastly better as well.

It has been suggested that Exploration might provide astronomy with cost offsets that would make lunar surface astronomy strategically attractive. In this picture, installation of a telescope at what might be a non-optimal site could have that non-optimality offset by not needing astronomical accounts to pay all the bills for it. Such cost offsets could, in principle, come from free transportation of equipment to the Moon (along with Exploration hardware), and ready infrastructure such as power plants and communication links. If full cost accounting does not apply to lunar surface astronomy, perhaps in an HST model in which vehicle development and operations costs do not come from an astronomy account, then the trade-offs could be more complicated. But astronomers have been given no assurance that such accounting manipulation is likely. As a result, and dismissing entirely one (non-science based) idea that an astronomical installation on the Moon could serve as a build-out focus for a human settlement, the strategic posture of the astronomical community has to be to look for ways in which the lunar surface might be *intrinsically* optimal for telescopes.

## 5. What Space Astronomy Needs from a Site

Evaluation of the lunar surface for astronomical instrument siting has to start with a list of requirements. Armed with such a list, astronomers are in a better position to compare the lunar surface with free space as an observatory site. One can develop such a list of requirements from the real mission concepts selected in decadal priority surveys (c.f. *Astronomy and Astrophysics in the New Millennium*, National Academy Press, Washington D.C., 2001) to meet established space astronomy needs. In brief, these requirements include

- precision optical alignment
- precision acquisition and tracking
- large field of regard
- low natural background
- large baselines and collecting areas
- low and stable temperatures
- quality (clean) optical surfaces
- assured communication and power
- opportunities for repair, routine service, and upgrade

It should be understood that these requirements are based on real mission concepts developed to meet science priorities, and that those were prioritized at least partly in the light of perceived capabilities. In this context, the last astronomy *Decadal Survey* panel did not consider the possibility of the lunar surface as a credible site for telescopes, and it can be (overgenerously, I believe) posited that their priorities might have been different if they did. It is also understood that these Decadal priorities were developed for the nearer term, while any kind of ambitious plans for lunar astronomy would likely have to take a much longer view. With this in mind, it is incumbent on the next Decadal survey to consider the unique characteristics of the lunar surface as they develop priorities and recommend technology investment and mission precursors.

## 6. Potential Problems for Lunar Surface Siting

The lunar surface as an observatory site can be considered with respect to the bulleted requirements listed above.

<u>Precision Optical Alignment</u> Precision alignment of optics is essential for realization of diffraction-limited performance. In the case of single-dish telescopes even the small (1/6g) gravity of the Moon is disadvantageous compared to free space in this regard. As the telescopes track across the sky, the changing gravity vector induces bending in the structure. While lunar surface telescopes could employ active linkages to correct for this bending (and with much lower bandwidth than needed on the Earth), or added weight for stiffer structures, this extra complexity and or weight will be a significant cost driver. Precision optical alignment puts strong requirements on thermal stability as well. In the case of TPF-C, for example, temperature stability at the 10mK level is needed. This kind of stability will not be realized on most of the lunar surface as the telescope tracks with respect to the surface, and the Sun tracks across the sky. Over the course of a lunar month, there will be large temperature changes in the structure, even if the telescope itself is surrounded by shielding.

<u>Precision Acquisition and Tracking</u> Although the lunar surface provides a slow-moving coordinate system compared with the Earth, the pointing and tracking requirements of large, diffraction-limited telescopes will be challenging. The performance of these telescopes will be limited both by gravitational bending modes (see above), but also by the quality of the bearing and drive mechanism. It has been pointed out that natural seismic activity on the Moon is quite low compared to the Earth (Mendell 1998), almost entirely dominated by meteoric impacts. But in the context of lunar development, it is likely that at least for telescopes convenient to the lunar support base, Exploration operations on the Moon (rovers, ascent/descent vehicles, mining operations) are likely to induce significant seismic noise. For the highest angular resolution telescopes that are envisioned, these effects need consideration.

Large Field of Regard  An unavoidable penalty of surface operations for a telescope is the $2\pi$ solid angle field of regard, and significantly less if the telescope is situated in a crater, as it might be at the lunar pole. In the case of an equatorial telescope, essentially the whole sky will be accessible, but not at one time. This may be disadvantageous for monitoring programs that might need information on timescales of weeks (e.g. supernovae and NEO tracking). Telescopes located near a lunar pole will be forever denied access to the other half of the sky.

Large Baselines and Collecting Areas  If we consider the Moon as an assembly platform for large telescopes, gravity may be disadvantageous. Delivery to the surface adds substantial risk and cost for every mass element. Furthermore, construction activities on the Moon depend upon lifting equipment. While it has been suggested that gravity is advantageous because constellation management of parts is a matter of just setting them down, rendezvous and docking of large components in free space is a proven capability. In the case of interferometric telescopes, the surface of the Moon might be considered a convenient and relatively stable optical bench. But the non-uniform surface of the Moon complicates optical linkage, and filling of the UV-plane involves moving interferometer elements around this non-uniform surface, most likely by a vehicle that can pick the elements up and set them back down, and perhaps using railcar tracks. Lessons learned from terrestrial spatial interferometers underscore the complex nature of repositioning and optically linking these surface-mounted individual elements.

Low Natural Background Emission  For infrared telescopes, shielding is challenging for a lunar surface mounted telescope, as (except for permanently shadowed lunar polar craters) the Sun and the Earth are difficult to block simultaneously. Permanently shadowed craters have estimated surface temperatures of order 30-60K and heat input is dominated by scattered and diffracted sunlight and to a lesser extent lunar heat flow. Such locations might be enabling in this respect for infrared background management, but those surface temperatures are not at all conducive to accessibility by astronauts and robots, and will require special measures to have power provided into them. While lunar dust is a challenge in many opto-mechanical respects, electrostatically levitated dust (see below) may present, by scattering sunlight, significantly elevated optical and ultraviolet sky background (Murphy and Vondrak 1993).

Low Temperatures  Following the discussion above, infrared telescopes on the Moon will benefit from low temperatures. Cosmic-background limited far infrared telescopes will need temperatures below 10K, and this is probably unachievable in the sunlit parts of the Moon. To the extent that permanently shadowed lunar polar craters are (as might be hoped for ISRU development) sites for frozen volatiles, volatilization by any ops-driven heat input, perhaps from mining activities, is not particularly conducive to contamination mitigation for a cold telescope, which will behave as an efficient cold trap for whatever evaporates are released.

Quality Optical Surfaces Contamination mitigation for optical surfaces is a critical need for high performance, both in terms of emissivity and scattered light. Control of the former is essential for background-limited infrared work, while control of the latter is essential for extrasolar planetary detection by optical and near infrared imaging. The tolerances for surface cleanliness are thus very tight. Naturally levitated dust, from both electrostatic and meteoric processes will need to be controlled, as will dust levitated by nearby surface operations. These issues are discussed in some detail below. The degradation of optical surfaces is just one facet of the many difficulties that lunar dust will present to equipment and astronauts on the Moon.

Assured Communications and Power Near-continuous solar power on the Moon is assured only in very limited areas (Malapert, etc.), and outside of these regions continuous supply of power will be a significant problem. While it is possible limit observatory operations to daylight, this both compromises the science productivity and perhaps adds significant risk to instrumentation. Direct communication with the Earth is only possible from the nearside of the Moon.

Upgrade and Repair Opportunities While having a telescope near a continually occupied lunar base would offer, in principle, the opportunity for repair and service, getting humans and their tools down safely to the lunar surface adds to mission risk and cost. In this age, with continuous human presence in free space and regular human trips to the Hubble Space Telescope, it is surprising that accessibility to space telescopes is often cited as an advantage somehow unique to the lunar surface. The nearsightedness of this notion will be elaborated upon below.

The challenge posed by lunar dust deserves special attention here, and has been considered by many authors in the context of lunar surface astronomy. One of the surprises of the Apollo program was how difficult lunar dust turned out to be. Gaier (2005) has recently reviewed these problems in detail with respect to the individual Apollo missions, and Johnson et al. (1991, 1995) have considered these difficulties in the context of lunar optical observatories. About a quarter by weight of the regolith material brought back by Apollo astronauts are particles smaller than 20μm, and they pose special hazards. The tiny, largely dielectric particles adhere electrostatically with great efficiency, and are also often sharp and shard-like such that they that hook and cling onto fabric. The low lunar gravity means that physical disturbances of any kind can disperse this dust on far-reaching ballistic trajectories. Dust deposits found by the Apollo 12 astronauts on Surveyor 3 were especially thick on the side facing the Lunar Exursion Module (LEM), which landed more than a hundred yards away. Apollo 17 astronauts needed to juryrig a fender on the lunar roving vehicle in order to suppress the dust that was being sprayed by it high over the surface. As documented by Gaier, the effects of dust degradation from the Apollo program were not just general dust coating, but also vision obscuration, false instrument readings, loss of foot traction, seal failures, clogging of mechanisms, material abrasion, thermal control problems, and inhalation and irritation risks. These were recognized on every lunar mission. Lunar dust particles carried back into the Apollo command module were found to be still problematical, even well off the lunar surface.

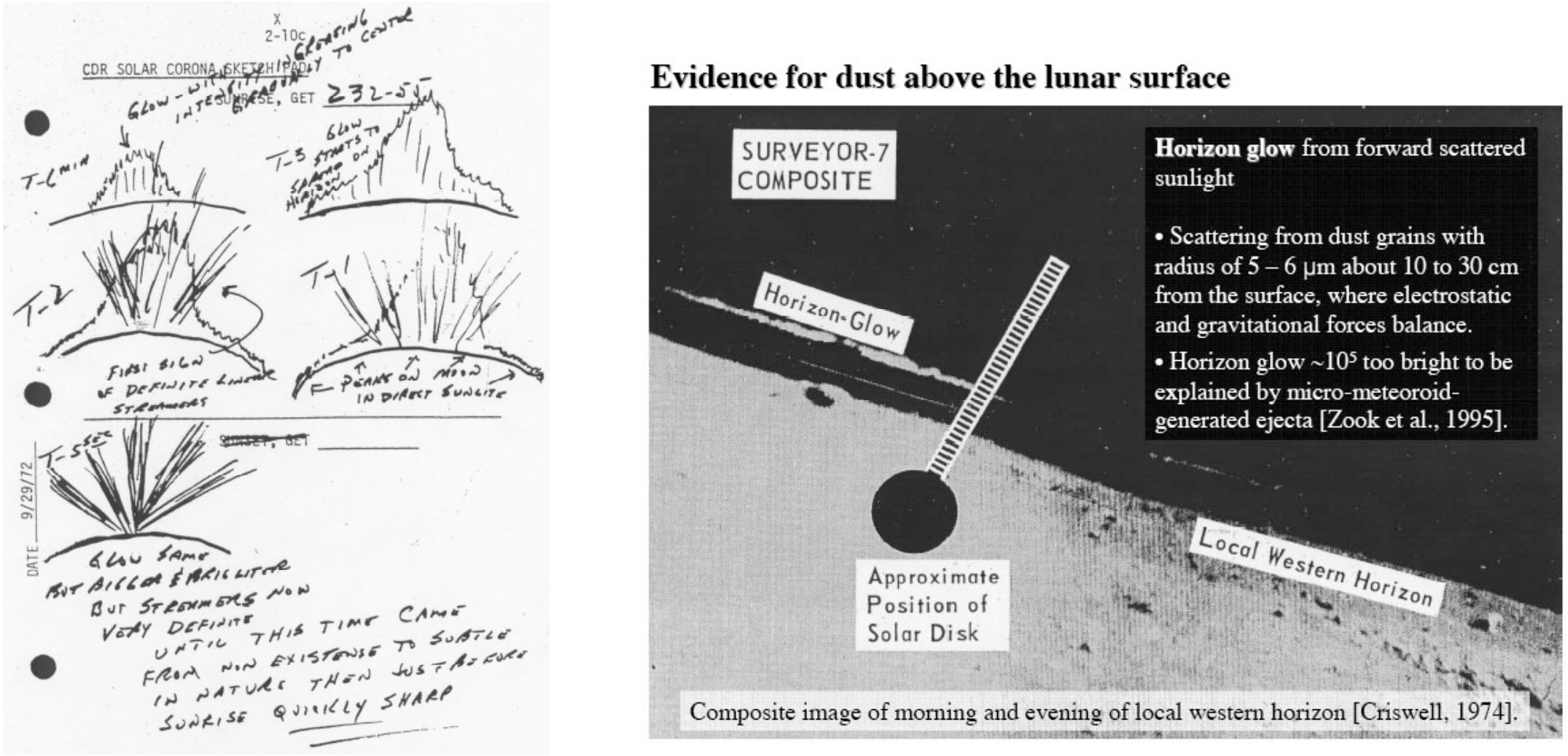

Figure 4: Naturally-levitated lunar dust is a well documented and remarkably conspicuous phenomenon, and poses a natural contamination risk to even remotely operated telescopes on the lunar surface. The dust is clearly visible in forward scattered light as what is termed lunar "horizon glow". This was seen visually at high elevation by Apollo 17 astronauts, the sketch at top left by Gene Cernan showing the appearance of the lunar horizon from the command module in low lunar orbit minutes before orbital sunrise. It was also easily detected under the same circumstances at low elevation along the local horizon by the Surveyor spacecraft (above right).

While surface operations-caused dust contamination was a serious issue for the Apollo program, the deposition of dust by natural processes is expected for future longer-term facilities, but the expected impact is not well quantified. Electrostatic dust levitation (originally proposed by Criswell 1972) is of particular concern in this regard. There is clear evidence for this naturally levitated dust above the lunar surface, and the evidence is not subtle. "Horizon glow" from forward scattered sunlight with the Sun just below the horizon was seen by both the primitive television cameras on the Surveyor 1,5,6 and 7 missions (Rennilson and Criswell 1974) with the naked eye by astronaut Gene Cernan in the Apollo 17 Command Module (McCoy and Criswell 1974), and likely later by the Clementine star tracker camera (Zook and Potter 1995). Evidence for deposition of this dust, ejected from the lunar surface by photoelectric charging of the tiny particles, comes directly from the Apollo 17 Lunar Ejecta and Micrometeorites (LEAM) experiment, which showed counts dominated by impacts from dust long after the LEM departure. As predicted by the latest electrostatic levitation models (Stubbs et al. 2006), the LEAM counts peaked near terminator passage (Berg et al.1974). So while the phenomenon of natural levitation of small particles is well established, and photoelectric lunar surface charging and a dynamic dust fountaining is the likely mechanism, column densities are not well known, and contamination rates that might be expected are thus uncertain.

It is noteworthy that surfaces regularly illuminated by sunlight are likely to be cleaned by this mechanism as well as contaminated (Stubbs, private communication). In this context, the Lunar Laser Ranging (LRR) reflectors left by the Apollo astronauts, and the condition of the Surveyor 3 spacecraft inspected by the Apollo 12 astronauts, may underestimate likely contamination of uplooking astronomical telescope mirrors well shielded from sunlight. While those LLR retroreflectors are still in regular use, their condition has not been carefully monitored by link efficiency calibrations over the last thirty-five years, at least in part because of the challenging photon return statistics (of order 0.01 photon per laser shot!) Recent careful attempts to quantify their reflectivity (Murphy 2006) with more than an order of magnitude better return statistics that are available with the new APOLLO system now suggest serious degradation, with one-pass retroreflector surface transmission only ~25% of that predicted. Such degradation is, fortunately, of minor relevance to the quality of the lunar laser ranging data. While there could be several causes of such degradation, naturally levitated dust deposited on the surfaces of the corner cubes is considered to be a likely culprit. Fragmentary historical accounts of link efficiency are being consulted to try to understand whether this degradation was not, in fact, the one-event plume of the LEM ascent. The consistency of the first returns with system predictions (E. Silverberg, private communication) suggests, however, that such one-time contamination was probably not the case. Although it would then appear that this retroreflector degradation has taken thirty five years to accumulate, it should be understood that even a few percent per year degradation by dust is extremely serious for astronomical optics that are used for thermal infrared work (in which emissivity, rather than reflectivity, is the limiting performance factor) and for high Strehl applications such as planet detection, in which scattered light must be minimized.

Figure 5: The lunar laser retroreflector packages (the one for Apollo 11 shown at right) offer the only available tests for long-term optical surface degradation on the Moon. Their performance has long been presumed to be undegraded, since they can still be used for ranging, but recent quantitative studies call this presumption into question. The optical contamination tolerances of modern astronomical telescopes is vastly lower than for the lunar laser ranging systems, and because these retroreflectors are not shielded from direct sunlight, dust accumulation on their surfaces may not be representative of that of astronomical telescopes, which are likely to be well shielded from direct sunlight.

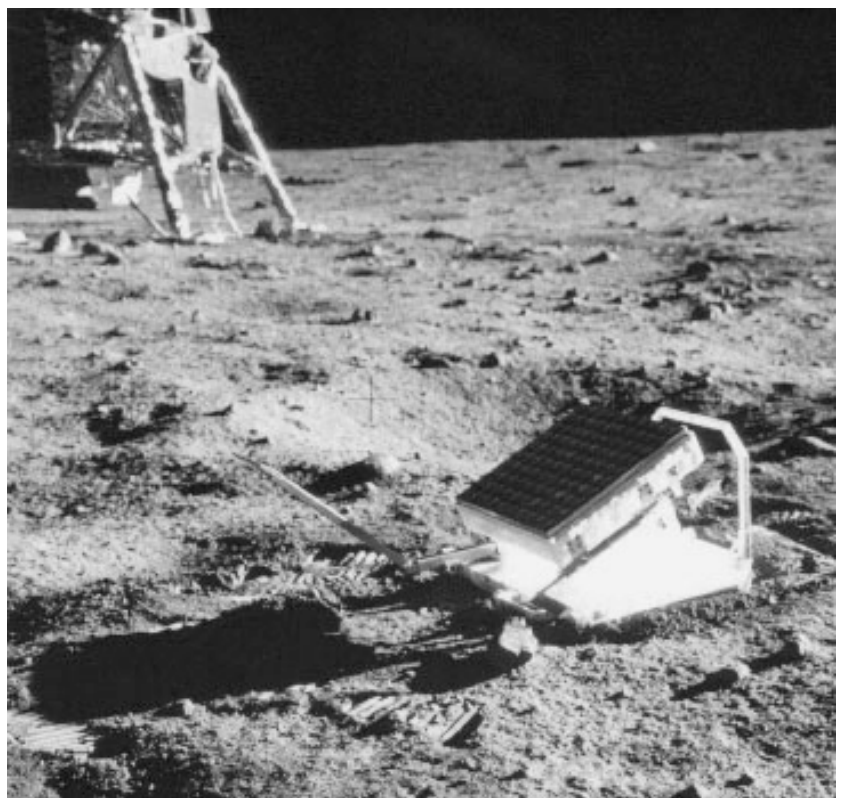

One of the primary needs for early term lunar exploration for both astronomy in particular and indeed for Exploration in general (c.f. Stubbs et al. 2005) is a quantitative assessment of the contamination efficiency of this natural rain of electrostatically lofted dust.

While several strategies for lunar dust mitigation have been proposed, it is clear that such strategies will add cost and inefficiencies to at least any short wavelength lunar surface astronomy. The threat of dust and mitigation strategies have not been the subject of this conference, so we have to defer to contamination engineers and lunar dust experts who will be considering the issues carefully. As we do so, it is important for the astronomical community to be reminded that the experiences the space program has had with lunar dust are daunting.

*"... one of the most aggravating, restricting facets of lunar surface exploration is the dust and its adherence to everything no matter what kind of material, whether it be skin, suit material, metal, no matter what it be and it's restrictive friction-like action to everything it gets on."*

*"There's got to be a point where the dust just overtakes you, and everything mechanical quits moving."*

*Gene Cernan; Apollo 17*

*"The LM was filthy dirty and it has so much dust and debris floating around in it that I took my helmet off and almost blinded myself. I immediately got my eyes full of junk, and I had to put my helmet back on. I told Al to leave his on." "We tried to vacuum clean each other down, which was a complete farce. In the first place, the vacuum didn't knock anything off that was already on the suits. It didn't suck up anything, but we went through the exercise."*

*Pete Conrad; Apollo 12*

*"Dust is the number <u>one</u> concern in returning to the Moon."*

*John Young; Apollo 16*

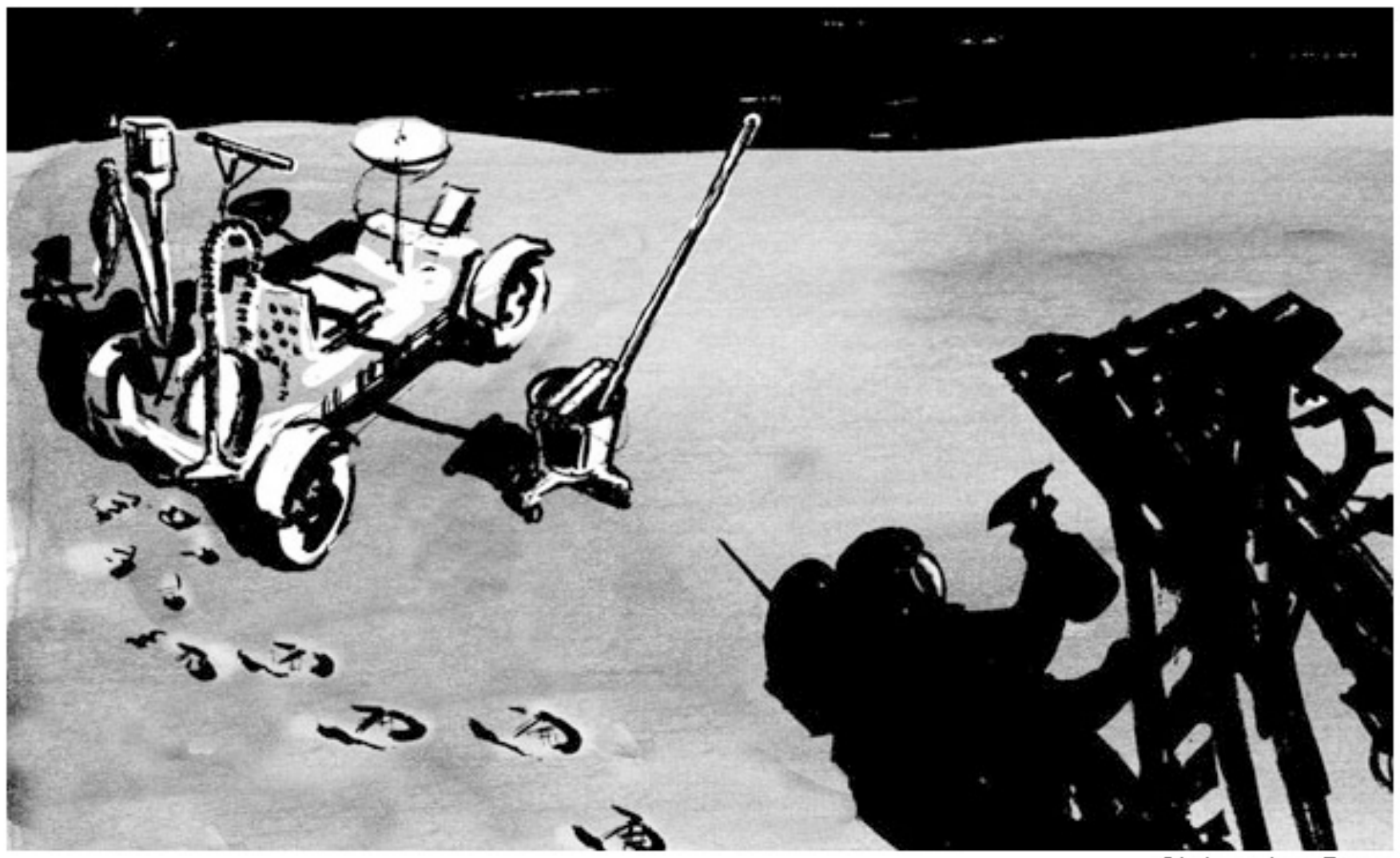

Figure 6: While various contamination mitigation strategies for lunar surface telescopes have been proposed, it can be hoped that techniques can be developed that actually keep dust off of lunar surface instruments, such that regular cleaning doesn't have to become part of standard maintenance.

## 7. The Advantages of Free Space

For most of the scientific priorities established by the community for space astronomy, free space can be argued to be a more productive and scientifically enabling site than the surface of the Moon. We do astronomy in free space right now, and we do it very well. Consideration of the value proposition for lunar surface astronomy thus has to take free space into account. The historical reluctance of lunar astronomy advocates to do this has been both conspicuous and surprising.

As for the surface of the Moon, the vacuum of free space offers essentially panchromatic operation of astronomical telescopes. The zero-g character of free space offers a large mass advantage for astronomical telescopes, which is reflected directly in development cost and deployment. With no gravity-induced stresses on the light collecting substrates or structure (though launch loads may be a relevant factor here), the telescope can be especially lightweight. As with the lunar surface, contamination from free-space propulsion systems and EVA suits needs to be managed carefully, though there is no dust or grit. Finally, at least for near Earth orbiting (e.g. LEO) telescopes, there is little latency in remote control operation from the surface of the Earth.

It has been understood that perhaps the most accessible and enabling place in free space for astronomy is the vicinity of the Earth-Sun 2$^{nd}$ Lagrange point (ES L2). This is a quasi-stable location, roughly four lunar distances from the Earth in the anti-Sun direction. Orbits around this location require some, but little stationkeeping, which is actually advantageous in that debris does not accumulate there naturally. WMAP is now operating successfully there, and ES L2 is the baseline site for the majority of future astronomy missions. The use of ES L2 for astronomical observations was originally proposed by Farquhar and Dunham (1990).

Earth-Sun L2 offers some remarkable enabling advantages for astronomy. With the Earth, Moon, and Sun largely in the same direction and with no periodic eclipses, satellites orbiting there can attain extraordinary thermal stability. With a modern multilayer solar radiation shield blocking these three heat sources, very low temperatures can be attained. JWST is designed in this way and, entirely passively cooled, is baselined to have an equilibrium temperature of <40K. Using the JWST design as a jumping off point, the Single Aperture Far Infrared (SAFIR) Vision Mission team has considered even more ambitious passive cooling, and strategies to passively reach temperatures below 15K there appear achievable (Lester et al. 2006). At ES L2, keeping a telescope cold passively just isn't that hard! Unlike magnetic and grav-grad torques in Earth orbit, E-S L2 is a remarkably force-free and low torque location, with the main dynamical perturbation being solar radiation pressure. The site has a continuous line-of-sight with Earth for communication and, with solar panels on the sunward side of any radiation shields, provides continuous, abundant solar power. Finally, as a roughly $C_3$=0 orbit, getting to Earth-Sun L2 is much easier propulsion-wise (by about 2.5 km/s in delta-V) than getting to the surface of the Moon, and without the risks of precision soft landing. In terms of thermal and dynamical stability as well as controllability, cleanliness, power availability, and modest distance from Earth, Earth-Sun L2 may truly be the ultimate site for astronomical telescopes.

So what does Earth-Sun L2 have to do with the return to the Moon and the *Vision for Space Exploration*? I believe that this ultimate site for astronomical telescopes can be strikingly served by the architecture being developed for lunar exploration. Certainly heavy-lift (Ares 5) launch technology could be called upon to put huge astronomical telescopes at ES L2. But the human and robotic priorities within VSE can be seen as profoundly enabling for space astronomy as well. While humans and robots on the lunar surface are usually argued as being potentially beneficial for space astronomy, it should be understood that such capabilities for hands-on deployment, servicing, upgrade and maintenance of astronomical facilities are already a regular feature of in-space operations. The Exploration Architecture will provide capabilities (e.g. CEV) that will extend our hands-on reach throughout cis-lunar space. For eventual travel to Mars, such in-space operational expertise will be of paramount important.

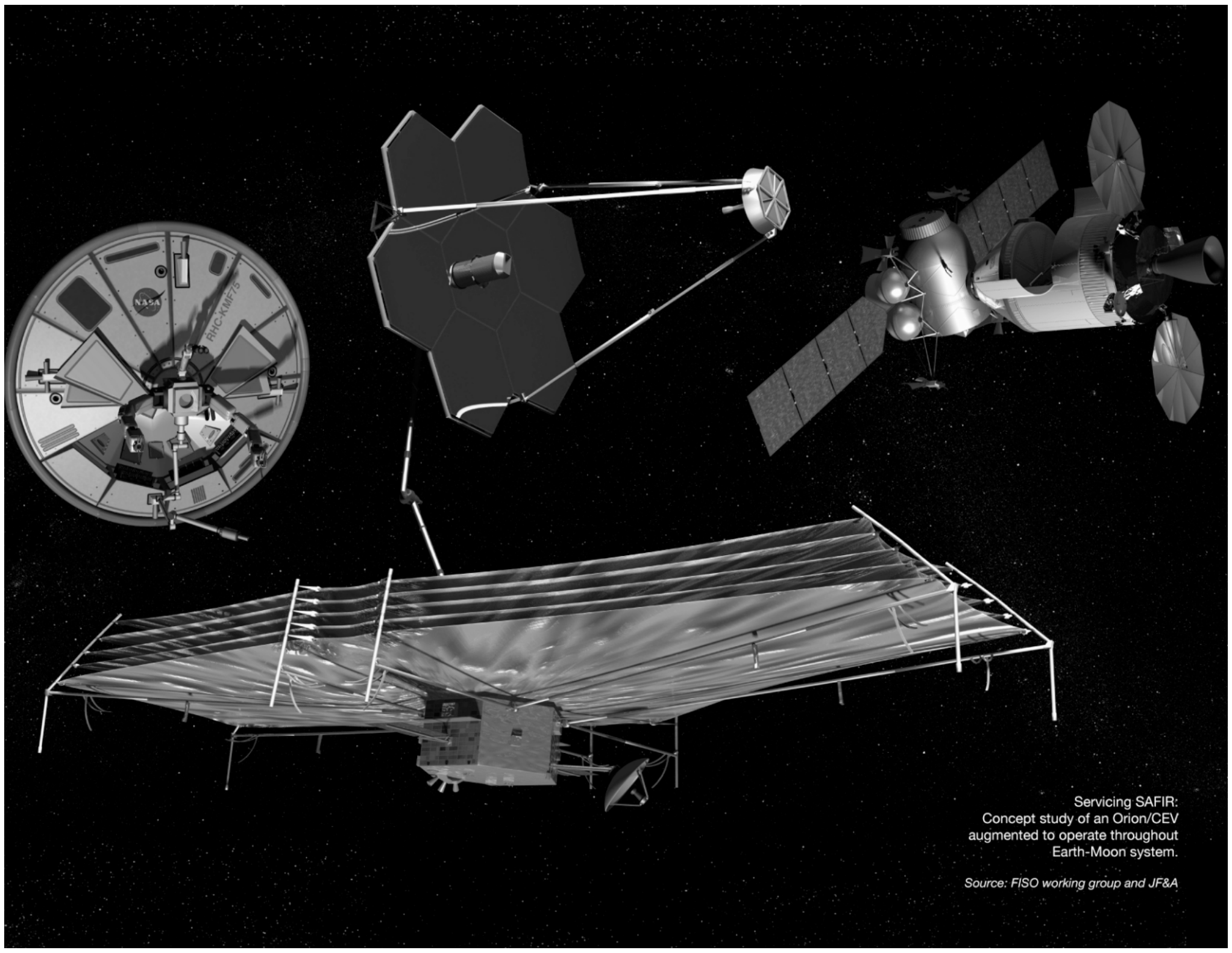

Figure 7: The Single Aperture Far Infrared (SAFIR) telescope is shown being serviced at Earth-Moon L1 by a CEV after having been returned to that location from its Earth-Sun L2 operational location on a low-energy pathway. A teleoperated robotic arm assists in the servicing effort, and is shown at far left. An LSAM module (perhaps one that had already been used for lunar surface operations, and stored at this location) is shown docked to the CEV, where it provides an airlock for astronaut EVA operations. Graphic from John Frassanito and Associates.

While human involvement in situ at ES L2 would probably challenge our capabilities at what we should consider the fringes of cis-lunar space, there are dynamical opportunities that we should take advantage of. It has been understood for many years that Lagrange points in the solar system are dynamically similar in potential energy, and that very low energy pathways connect them along routes that may well be quite indirect (Lo and Ross 2001, Ross 2006). In this context, the Earth-Moon L1 and L2 Lagrange pointswhich are 15% of the Earth-Moon distance on the near- and far-side of the Moon, have special appeal. While these locations do not offer the thermal isolation of Earth-Sun L2, they offer the convenience of being easily accessible to the planned lunar Exploration architecture. Getting humans and robots to these locations is easier than getting them to the lunar surface. Getting our astronomy missions back and forth between ES L2 – the ultimate operational site for astronomical telescopes – and EM L1 and L2 is simple, requiring a delta-V of merely tens of meters per second (and a somewhat leisurely transit time of several months). Such propulsion loads can be easily borne by such space observatories, producing little stresses on potentially fragile structures. Infrared telescopes that cool passively can be conveniently warmed for service calls simply by rotating them so sunlight can hit them, unlike for lunar polar crater telescopes. The Earth-Moon Lagrange points then become potential jobsites for astronomical observatory deployment and servicing. This opportunity has been recently been considered for SAFIR servicing (Lester, Friedman, and Lillie 2005), and also for more general cases (Stevens and King 2005). For a broad view, the reader is referred to Harley Thronson's contribution to this workshop – "*Adapting NASA's Exploration Architecture to Achieve Major Astronomy Goals in Free Space*."

## 8. The Lunar Surface as an Enabling Site for Astronomical Observations?

In view of the potential difficulties with using the lunar surface for astronomy, and the advantages of free space, it is essential that science communities look hard at the unique advantages that the lunar surface might well offer. In this context we return to the point expressed above – that the Moon uniquely offers, with no atmosphere and relative proximity, *rocks* and *gravity*. What can we do with those? This workshop has featured a number of clever ideas for using rocks and gravity to benefit astronomy, and these should be considered carefully with regard to both technical feasibility and scientific priority.

One exciting idea reviewed in the workshop is the use of the Moon as a shield against terrestrial radio interference (both from human-operated transmitters on the ground and in GEO, as well as natural radiation from the geomagnetic auroral zone). The Radio Astronomy Explorer satellite (RAE-2) was launched into an inclined lunar orbit in 1973, with 13-25 MHz receivers fed by large ~200m long V-dipole antennae (Alexander et al. 1975). The satellite was in a 1000 km high orbit, low enough that the Earth and Sun were occulted by the Moon, which subtended a disk size of ~76°. RAE-2 showed remarkable drops in the ambient radio power density during each such occultation – by almost two orders of magnitude for an Earth occultation, and less for an occultation of the Sun. RAE-2 thus established that the lunar farside is, by virtue of it being a large rocky body, the quietest radio location in the Earth-Moon system. Such galactic background-limited performance is not achievable anywhere else nearby, and can be hugely enabling for low frequency radio cosmology probes. Terrestrial implementations of low frequency radio interferometers (e.g. SKA, LOFAR) are designed to address high priority astronomical questions, but are limited by their terrestrial siting, and the enormous challenge of interference rejection. Using the farside of the Moon for such a telescope is thus of significant interest, and several participants of this workshop, including Jackie Hewitt and Chris Carilli, have presented ideas on it. It should be noted that telescopes down at ground level, perhaps inside craters where more than half the sky in the direction of the Earth is blocked, are likely to be even better shielded from terrestrial interference than was RAE-2.

There are trade studies that need to be undertaken for this lunar shielding option, however. Firstly, the extent to which lunar farside siting to minimize terrestrial interference is actually preferable to a free-space siting option in which a telescope is simply sent to a large distance from the Earth. It should be noted in this context that free space radio interferometry is proven technology (e.g. HALCA/MUSES-B). Secondly, the protection of the"quiet zone of the Moon" (QZM) as it is referred to, is of great importance in this regard. The development of such a telescope by a farside-capable space program may well involve pollution of the radio environment by the concomitant local transmitters and relay stations. Finally, development on the far side of the Moon is not an obvious element of near-term lunar exploration. Do we have good reasons for farside development? If not, then a farside telescope would not benefit from early infrastructure there. Finally, in order to proceed with planning for such a facility, there is a clear need for site surveys that would better characterize the radio background over long timescales, with special consideration to passage of the Moon through the Earth's geotail and solar activity.

Another idea of interest is using the gravity of the Moon to form large parabolic telescope mirrors made of spinning liquid. Roger Angel and Ermanno Borra have led studies of this creative idea, and discuss it here. In this case, the lunar gravity is an enabling characteristic of the site. In principle, very large collecting areas can be "assembled" with a small volume of liquid. The technological feasibility of such a telescope, which might have an aperture size of 20-100m, is a huge challenge however, depending on a large mechanical installation with precision bearings. In order for this telescope to have high performance in the infrared, it would probably be built where the Sun could be well shielded, probably at a pole, and an active liquid would need to be chosen that would not freeze, and yet could support a flashed-on reflective layer. Such a telescope would view the local zenith only, and high priority science drivers for deep, small-field operation would have to be agreed upon. Such a telescope might be particularly sensitive to dust contamination, as surface cleaning would be difficult. Although a liquid parabolic mirror could not be formed in this way in free space, a valid trade study would compare the cost and feasibility of this telescope to a more conventional mega-telescope in free space, which would be freely pointable.

Finally, the idea of using the lunar surface as an optical bench for optical and infrared interferometers has attracted attention. In this picture, the rocky body of the Moon provides a surface that keeps interferometer elements from moving around, obviating the stationkeeping and formation-flying requirements that would be needed for free space.

While the lunar surface is stable, it is not smooth, and distributed telescopes would have to be linked across rocks, boulders, and larger landforms. Unlike free space deployment, such a lunar-based telescope makes redeployment of collectors difficult. In order to fill the UV plane, one has to pick them up and move them around. While a fixed surface offers some advantages for baseline management, the impact of large monthly temperature swings and power availability would need careful consideration. Jack Burns has presented some thoughts on this in this workshop, injcuding a novel idea for a telescope array embedded in an unrollable plastic sheet. While our space program does not presently have the technology for such short wavelength interferometers in free space, studies of LISA, and in the longer term, TPI-I, DARWIN, and SPECS suggest that extrapolation to the necessary technology is entirely credible, leading to telescopes that, because they could be set up in a wide range of configurations, could be more scientifically productive.

Finally, several contributors have considered using the rocky body of the Moon as a cosmic high-energy radiation and particle telescope. In this case, in which a large mass of the Moon is used as a detector, there are probably no free space options that would compete!

## 9. Exploration and Astronomy Doesn't Have to Mean the Lunar Surface

Through this workshop, the astronomical community has been challenged to carefully consider astrophysics that would be enabled by a return to the Moon. As we do this, we must resist the temptation to reflexively assume that such a return to the Moon can only offer us telescopes that are "down on the rocks." Similarly, the Exploration community must resist a temptation to reflexively presume that, just because astronomers once saw their science as being broadly enabled by lunar surface telescopes, it still is. A careful assessment of the value proposition for astronomy in the Exploration era will include human capabilities we now have in free space and build upon the vast experience in construction, maintenance, and servicing we have developed through ISS as well as through astronomically focused operations with HST.

## Acknowledgements

In formal reconsideration of the lunar surface as a site for astronomy, I'd like to express my appreciation to John Mather, Hal Yorke, and Harley Thronson for their careful perspectives on relevant trade studies, and to the Future In-Space Operations Working Group and the SAFIR Vision Mission team for helping to conceptualize the role of Exploration architecture in serving astronomical needs. As a contributing member of the NRC panel on the Scientific Context for the Exploration of the Moon, it should be understood that the opinions expressed here are not intended to represent the conclusions of that panel.